\documentstyle[referee]{mn}

\def\beq{\begin{equation}}
\def\eeq{\end{equation}}
\def\bey{\begin{eqnarray}}
\def\eey{\end{eqnarray}}
\def\RM{\rm}
\input epsf

\title[]
	{On the Microlensing Optical Depth of the Galactic Bar}
\author[]
	{HongSheng Zhao and Shude Mao
	\thanks{E-mail: (hsz, smao)@mpa-garching.mpg.de} \\
	Max-Planck-Institute f\"ur Astrophysik,
Karl-Schwarzschild-Strasse 1, 85740 Garching, Germany}
\date{Accepted ........
      Received .......;
      in original form .......}
\pagerange{\pageref{firstpage}--\pageref{lastpage}}
\pubyear{1996}

\begin{document}
\maketitle
\label{firstpage}

\begin{abstract}

The microlensing probability (optical depth $\tau$) toward the
Galactic center carries information about the mass distribution of the
Galactic bulge/bar, so can be used to constrain the very uncertain
shape parameters of the Galactic bar.  In this paper we study a family
of bar models with tunable triaxiality, boxyness and radial profile to
span the range of plausible density models of the Galactic bar.  It
also includes four models that were used to fit to the COBE maps by
Dwek et al (1995).  We find the optical depth on the minor axis of the
whole family of models and in special cases the whole microlensing map
of the bulge/bar are given by simple analytical formulae.  The
formulae show the following dependence of the optical depth on the bar
mass, radial profile, angle, axis scale lengths and boxyness.

(1) $\tau$ is proportional to the mass of the bar, $M$.

(2) $\tau$ falls along the minor axis with a gradient determined by
the density profile and boxyness of the bar.

(3) Going from an oblate bulge to a triaxial bar, the optical depth
increases only if $\alpha<45^\circ$, where $\alpha$ is the angle between
the bar's major axis and our line of sight to the center.

(4) Among bars with different degree of triaxiality, $\tau$ is the
largest when the bar axis ratio ${y_0 \over x_0} =\tan \alpha$, where
$y_0, x_0$ are the bar scale lengthes on the short and long axes in
the Galactic plane respectively.

(5) At a fixed field on the minor axis but away from the center, boxy
bars with a flat density profile tend to give a larger optical depth
than ellipsoidal bars with a steep profile.

(6) Main sequence sources should have
 a significantly lower (20-50\% lower) optical depth than red
clump giants if main sequence stars are not observed as deep
as the bright clump giants.

For the four COBE-constrained models, the optical depths contributed
by the bar is in the range of $\approx (0.3-2.2) \times 10^{-6} M/(2
\times 10^{10} M_\odot)$ at Baade window. Out of the four models,
only the model with a Gaussian profile and ellipsoidal shape satisfies
the observational constraint, the other three models produce optical
depths which are inconsistent or marginally consistent with the
$2\sigma$ lower limit of the observed optical depths, even if we adopt
both a massive bar $2.8\times 10^{10} M_\odot$ and a full
disk. Independent of the COBE map, we find that the microlensing
models can potentially provide constraints on the mass
distribution. If increasing microlensing statistics confirm the high
optical depth, $\approx 3\times 10^{-6}$, presently observed by the
MACHO and OGLE collaborations, then the observation argues for a
massive ($\ge 2\times 10^{10}M_\odot$) boxy bar with axis ratio 
${y_0 \over x_0} \approx \tan(\alpha)$ and $\alpha \le 20^\circ$ and
with a flat radial profile up to corotation.  Stronger limits on the
bar parameters can be derived within two years when the sample is
increased from the current fifty events to about two hundred events.

\end{abstract}


\section{Introduction}

Microlensing surveys such as the MACHO (Alcock et al. 1993, 1995a, b),
OGLE (Udalski et al. 1993, 1994), EROS (Aubourg et al. 1993) and DUO
(Alard et al. 1995) collaborations have discovered more than
100 microlensing events toward the Galactic bulge and about 10 events
toward the LMC.  One of the many exciting discoveries (see Paczy\'nski
1996 for a review) is the surprisingly high optical depth, $\tau$, i.e.,
microlensing probability, toward the Galactic centre (Udalski et
al. 1994; Alcock et al. 1995b). The OGLE collaboration found $\tau_{-6}
\equiv \tau/10^{-6}=(3.3 \pm 2.4)$ ($2\sigma$ error bar)
based on 9 events and the MACHO collaboration found
$\tau_{-6}=2.43 ^{+0.9}_{-0.74}$ ($2\sigma$) based on the full sample
of 41 events and $\tau_{-6} = 6.32^{+6}_{-3.57}$ ($2\sigma$) for 10
low latitude clump giants.  This is clearly in excess of all the
predictions prior to the discovery of microlensing events (Kiraga \&
Paczy\'nski 1994a). Both a maximum disk (Gould 1994a, Alcock et
al. 1995b) and a Galactic bar oriented roughly toward us (Paczy\'nski
et al. 1994a; Zhao, Spergel \& Rich 1995, Zhao, Rich \& Spergel 1996,
hereafter ZSR95, ZRS96) have been proposed as an explanation.
The latter interpretation is favored by other lines of evidence, such
as the non-circular motion of gas in the inner Galaxy
(Binney et al. 1991; Blitz \&
Spergel 1991), the asymmetric and boxy COBE map (Weiland et al. 1995) and
that the clump giants appear brighter at one side ($l>0^\circ$) than
the other side ($l<0^\circ$) of the bulge (Stanek et al. 1994).
A Galactic bar, together with a normal disk, can produce an optical
depth $\tau_{-6} \approx 2.5$ (ZSR95, ZRS96). This value is in good
agreement with the OGLE and the MACHO full sample values, but seems to
be at the $2\sigma$ lower limit of that for the 10 low latitude clump
giants in the MACHO sample.
%
The theoretical prediction depends on the still uncertain bar axial
ratio, orientation and mass profile; Dwek et al. (1995) found that
several bar volume density models are consistent with the COBE map.
Given these uncertainties, what
is the range of the predicted optical depth? What are the optimal bar
configurations that can produce an optical depth as high as
$\tau_{-6}=6$? Can the Galactic microlensing observations put constraints
on the bar parameters? These are the main questions we want to address in the
paper.
%
We demonstrate that the published optical depth based on about 10
microlensing events with red clump giants as sources already put
interesting constraints on the orientation, axial ratio and boxyness
of the bar. With steadily increasing number of events and decreasing
error bar for the optical depth, microlensing will become a very
powerful tool to probe the Galactic structure.

The outline of the paper is as follows. In \S 2, we study a family of
models for the bar density distribution, which includes many models
used in the literature. We derive an analytical expression of their
optical depths on the minor axis and for ellipsoidal bars the whole
microlensing map on the sky as well. In \S 3, we apply our formalism
to the Galactic bar, and compare the results with the observations.
We summarize our results and discuss the implications in \S 4. More
mathematical details are given in the appendices and at the end of
Appendix C some instructions of using our formulae.

\section{Model}

In this section we first give the general expressions for the
microlensing probability (optical depth) of a bar, then derive 
the optical depth map for a simple ellipsoidal
Gaussian bar model. The results are then extended to 
a family of bar models with more general shapes and radial profiles.  
Using these models we study the dependence of the optical depth on 
parameters of the bar.

\subsection{Microlensing Optical Depth of the Bar}

The microlensing optical depth for a source at distance $D_s$
is the probability for it to come within one Einstein radius of
any lens (deflector) placed at $0 < D_d < D_s$:
\beq
\tau(D_s)=\int_0^{D_s} \left({\rho(D_d) \over m}\right) 
                 (\pi R_E^2)~ dD_d, \,\,
		 R_E \equiv \left( {4 G m D \over c^2} \right)^{1/2},~
		 D \equiv (D_s-D_d) {D_d \over D_s},
\eeq
where $m$ is the mass of the lens, $\rho(D_d)$ is the mass density of
the lens at distance $D_d$, $R_E$ is the Einstein radius,
$D$ is the effective distance between the lens and the source,
$c$ is the speed of light and $G$ is the gravitational constant.
The first bracketed term is the number density
of the source at distance $D_d$, while the second term gives the 
lensing cross section, namely, the area enclosed by one Einstein radius.
Note that the lens mass $m$ cancels out, therefore the optical depth
depends only on the total mass density of the lenses,
but not on the mass spectrum.

The microlensing optical depth is usually measured for all sources
in a narrow cone of the sky 
(e.g., Baade window of the bulge, $l=1^\circ, b=-3.9^\circ$),
therefore it is necessary to take into account the distribution of 
source distances by averaging over the line of sight:
\beq \label{tauav}
\langle \tau \rangle= {\int_0^{s_{max}} \tau(D_s) n(D_s)dD_s \over \int_0^{s_{max}}  n(D_s)dD_s },\,\, 
n(D_s)~dD_s \propto \rho(D_s) D_s^{2-2 \beta_s}~dD_s,
\eeq
where $D_{s,\mathrm{max}}$
is the maximum distance of an observable source, and
$n(D_s) dD_s$ is the number of observable sources 
at distance $D_s$ to $D_s+dD_s$ in the field of view.
The factor of $D_s^2$ in $n(D_s)$ takes into account the volume increase
with distance for a fixed field of view, and the factor $D_s^{-2 \beta_s}$
parametrizes the decrease of detectable sources with distance due to
the detection threshold (Kiraga \& Paczy\'nski 1994). 
For the main sequence stars, Kiraga \& Paczy\'nski (1994) and ZSR95
suggested $1 \le \beta_s \le 2$. For the clump giants,
$\beta_s=0$ is more approriate as they are probably seen throughout
the Galaxy. Note that $\langle \tau \rangle$ is generally a function of 
Galactic coordinates, but we ignore the angular variation
inside a small field of view (less than one square degrees).

In all the following sections, we will use the Galacto-centric
coordinate system $(x, y, z)$, with the Galactic centre at the origin,
and the solar system at $(-R_0, 0, 0)$, where $R_0=8.5$ kpc is the distance to
the Galactic centre.  $(X,Y,Z)$ are used to denote the coordinates
along the three principal axes of the bar; $X$ being the major axis of the
bar.  Since the tilt out of the Galactic plane is negligible 
(Dwek et al. 1995), the $(X,Y,Z)$
coordinate system is simply related to $(x, y, z)$ by 
a rotation of angle $\alpha$ in the $x-y$ plane.
It is understood that only the absolute value of the angle,
$|\alpha|$, is used.

Since the longest dimension of the bar, $a$, is still much smaller
than the bar's distance from us, $R_0$, we can make two
simplifications to eq. (\ref{tauav}) so that the calculations are
tractable analytically:

(1) We set $n(D_s) = \rho(D_s)$, i.e., $\beta_s=1$, so that the volume
increase with distance just balances the decrease of detectable
sources. This is a plausible assumption for the bulge main sequence
stars, but probably underestimates the optical depth for the clump
stars with $\beta_s=0$ by an order of $({a \over R_0})$.

(2) We adopt a plane parallel approximation:
$$
x_s \approx D_s-R_0, \,\,\, x_l \approx D_d-R_0, \,\,\, {D_d \over D_s} \approx 1, \,\,\, D = (D_s-D_d){D_d \over D_s}
\approx x_s-x_l,\,\,\,	
$$
and the Galactic longitude and latitude of any bulge star
$$
(l, b) \approx ({y \over R_0}, {z \over R_0}).
$$
Here $(y,z)$ are the coordinates
where the line of sight intersects with $y-z$ plane at the center. 
$x_s$ and $x_l$ denote
the $x$-coordinates of the source and the lens, which 
are assumed to be in the range of
$$
-\infty <x_l \le x_s<+\infty.
$$
In making this simplification, we effectively 
set $-R_0 \rightarrow -\infty$, and $s_{max} \rightarrow
\infty$.  This is valid for lenses in the bar
as long as the density of the bar
falls off sufficiently fast or is truncated within a few kpc,
which is the case for all our models. Note
by letting ${D_d \over D_s}=1$ we
overestimate $D$, and hence the optical depth, by an
order of $({a \over R_0})$.  The assumption also
neglects the difference in distance between the
positive and negative longitude sides.

With these simplifications, the optical depth for a source located
at $(x_s, y, z)$ is given by (cf. eq. 1)
\beq\label{taus}
\tau(x_s,y,z) = {4 \pi G \over c^2}  \int_{-\infty}^{x_s} \rho(x_l, y,z) ~ (x_s-x_l)~ dx_l.
\eeq
while the average optical depth toward a field $(y, z)$ is (cf. eq. 2)
\beq	\label{taua}
\left< \tau(y,z) \right> = {\int_{-\infty}^{+\infty}dx_s ~ \rho(x_s, y, z) ~
\tau(x_s,y,z) \over \int_{-\infty}^{+\infty}dx_s ~ \rho(x_s, y, z) }.
\eeq

For the clump giants, the effects of our assumptions (1) and (2)
on $\tau$ are in the opposite direction. Remarkably, numerical
calculations find that these two effects cancel out,
therefore, our analytical results based on eqs. (\ref{taus}) and
(\ref{taua}) are accurate for the clump giants (cf. Fig. 3.)  
For the fainter main sequence stars eq. (\ref{taua}) overestimates the
optical depth.  But this can be easily corrected analytically to the
first order of ${ a \over R_0}$ using 
eqs. (\ref{xi4}) and (\ref{correct}) in Appendix C.
The corrected values are again in good agreement with numerical
results. The analytical formulae, of course, have the
significant advantage that they allow the
dependence on the parameters to be studied explicitly (see \S 2.3.)

\subsection{Analytical Results For a Gaussian Ellipsoidal Bar}

Consider the optical depth of the Galactic bar 
with a Gaussian radial profile and an ellipsoidal shape,
which is one of the models that Dwek et al. (1995)
used to fit the COBE map.  The density of this model is given by
\beq
\rho(x,y,z)=\rho_0 ~ \exp \left(-{\lambda^2 \over 2}\right), ~ 
\rho_0= {M \over (2 \pi)^{3 \over 2} x_0 y_0 z_0} ,
\eeq
where
\beq\label{reff}
\lambda^2= \left({X \over x_0}\right)^2 + \left({Y \over y_0}\right)^2 + \left({Z \over z_0}\right)^2,
\eeq
$M$ is the total mass of the bar, $x_0, y_0, z_0$ are the scale lengths along
the three principal $(X,Y,Z)$ axes of the bar and $\rho_0$
is the central density. $X,Y,Z$ are related to
$x,y,z$ by a rotation of the bar angle $\alpha$,
\beq \label{transformation}
X  =  x \cos \alpha - y \sin \alpha,~~
Y  =  x \sin \alpha + y \cos \alpha,~~
Z  =  z.
\eeq
Substituting eq. (\ref{transformation}) into eq.~(\ref{reff}), one finds that
\beq
\lambda^2= \left({x \over x_1}\right)^2 + \left({y \over y_1}\right)^2 + \left({z \over z_1}\right)^2 + 2 \kappa xy,
\eeq
where $x_1$, $y_1$ and $z_1$ are effectively the scale lengthes along
the $x$, $y$ and $z$ axes.  Together with $\kappa$ they are given by
\beq\label{x1y1z1}
{1 \over x_1^2}   =  {\cos^2 \alpha \over x_0^2} + {\sin^2 \alpha\over y_0^2},~~
{1 \over y_1^2}   =  {\sin^2 \alpha \over x_0^2} + {\cos^2 \alpha\over y_0^2},~~
{1 \over z_1^2}   =  {1 \over z_0^2},~~
\kappa               =  ({1 \over y_0^2}-{1 \over x_0^2})
\sin \alpha \cos\alpha.
\eeq

So for the Gaussian ellipsoidal bar, the optical depth for a source 
at the Galactic 
centre is (cf. eq. \ref{taus})
\beq
\tau(0,0,0)  = {4 \pi G \over c^2}
\int_{-\infty}^0 \rho_0 \exp (- { x_l^2 \over 2 x_1^2} ) (-x_l) dx_l
             =  {4 \pi G \over c^2}  \rho_0 x_1^2.
\eeq
The optical depth for a source on the minor axis at $(0, 0, z)$ is given by
\beq
\tau(0,0,z) = \tau(0,0,0) \exp (- { z^2 \over 2 z_0^2}).
\eeq
The optical depth for all sources along a line of sight passing the minor axis
can be obtained from eq. (\ref{taua})
\beq \label{taugauss}
\left< \tau(0,z) \right>  =  \sqrt{2} \tau(0,0,z)
			  = {4 \pi G \over c^2}  {M \over z_0}
			  {1 \over \pi^{1/2}}
			  I({y_0 \over x_0},\alpha)
\exp (- { z^2 \over 2 z_0^2} ),
\eeq
where
\beq	\label{IG}
I({y_0 \over x_0},\alpha)
={ 1 \over 2\pi}~
\left[\left({y_0 \over x_0} \cos^2\alpha\right) + 
\left({x_0 \over y_0} \sin^2\alpha\right) \right]^{-1}
= {1 \over \pi} ~
{\left( u+u^{-1} \right)^{-1} \over \sin 2 \alpha}, 
~~~~ u \equiv{ {y_0 / x_0} \over \tan \alpha}.~
\eeq
Similarly we can obtain a microlensing optical depth 
map on the whole sky, i.e., on
the whole $y-z$ plane:
\beq \label{gausstaumap}
\left< \tau(y,z) \right>  = 
{4 \pi G \over c^2} {M \over z_0} {1 \over \pi^{1/2}} I({y_0 \over x_0},\alpha)
\exp (- { y^2 \over 2 {y^\prime_1}^2} - { z^2 \over 2 z_0^2} ),~~~~
y^\prime_1 \equiv \sqrt{x_0^2 \sin^2 \alpha
+y_0^2 \cos^2 \alpha}
\eeq
Note that for any field with the same $z$, the optical depth is the
largest on the minor axis.  

\subsection{Analytical Results For a Family of Bar Models}

The bar density is not uniquely constrained by the COBE map 
(Dwek et al. 1995).
Models with various radial profiles and shapes can fit the data about
equally well.  In this section, we derive the optical depth
for a general set of bar density models and show its
dependence on the profile and shape of the bar.

Consider the following class of bar density models:
\beq \label{rhoxyz}
\rho(x,y,z)=\rho_0 f(\lambda), ~~~
f(\lambda) \equiv \left(1+ {\lambda^n \over \beta-3}\right)^{-{\beta \over n}},
\eeq
where the central density $\rho_0$ can be expressed in terms of
the total mass of the bar, $M$, and is given in eq. (A1) in the Appendix A.
The effective dimensionless radius parameter $\lambda$ is given by
\beq \label{lambdan}
\lambda^n= \left[\left|{X \over x_0}\right|^k + \left|{Y \over y_0}\right|^k
\right]^{n \over k} + \left|{Z \over z_0}\right|^n.
\eeq
The parameters $n$ and $k$ determine the shape of the bar:
$k$ specifies the shape of the bar in the $X-Y$ plane
while $n$ specifies the shape in the $Y-Z$ and $X-Z$ planes. The
parameter $\beta$ describes the radial profile of the bar.
When $\beta>0$, the density falls off like a power-law with slope $(-\beta)$
at large radius. When $\beta \le 0$, the model describes
a truncated finite bar confined within $\lambda \le |\beta-3|^{1 \over n}$.
When $\beta \rightarrow \infty$, one
obtains an exponential or Gaussian
density profile for $n=1$ or $n=2$. The Gaussian ellipsoidal model 
studied in the previous section corresponds 
to a special case with $(\beta, n, k)=(\infty, 2, 2)$.
A homogeneous rectangular bar corresponds to a model with $(\beta, n,
k)=(0, \infty, \infty)$; analytical results for this special case
are given in the Appendix B.

This family embodies the more interesting cases of bar models\footnote{
models with $0<\beta\le 3$ correspond to flaring bars with 
their densities increase 
with radius, and will not be considered further.} and can simulate
a great diversity of shapes and radial profiles.
The range of shapes and profiles are demonstrated by 
a few examples in Fig. 1a and 1b.

\begin{figure}
\epsfysize=10cm
\centerline{\epsfbox{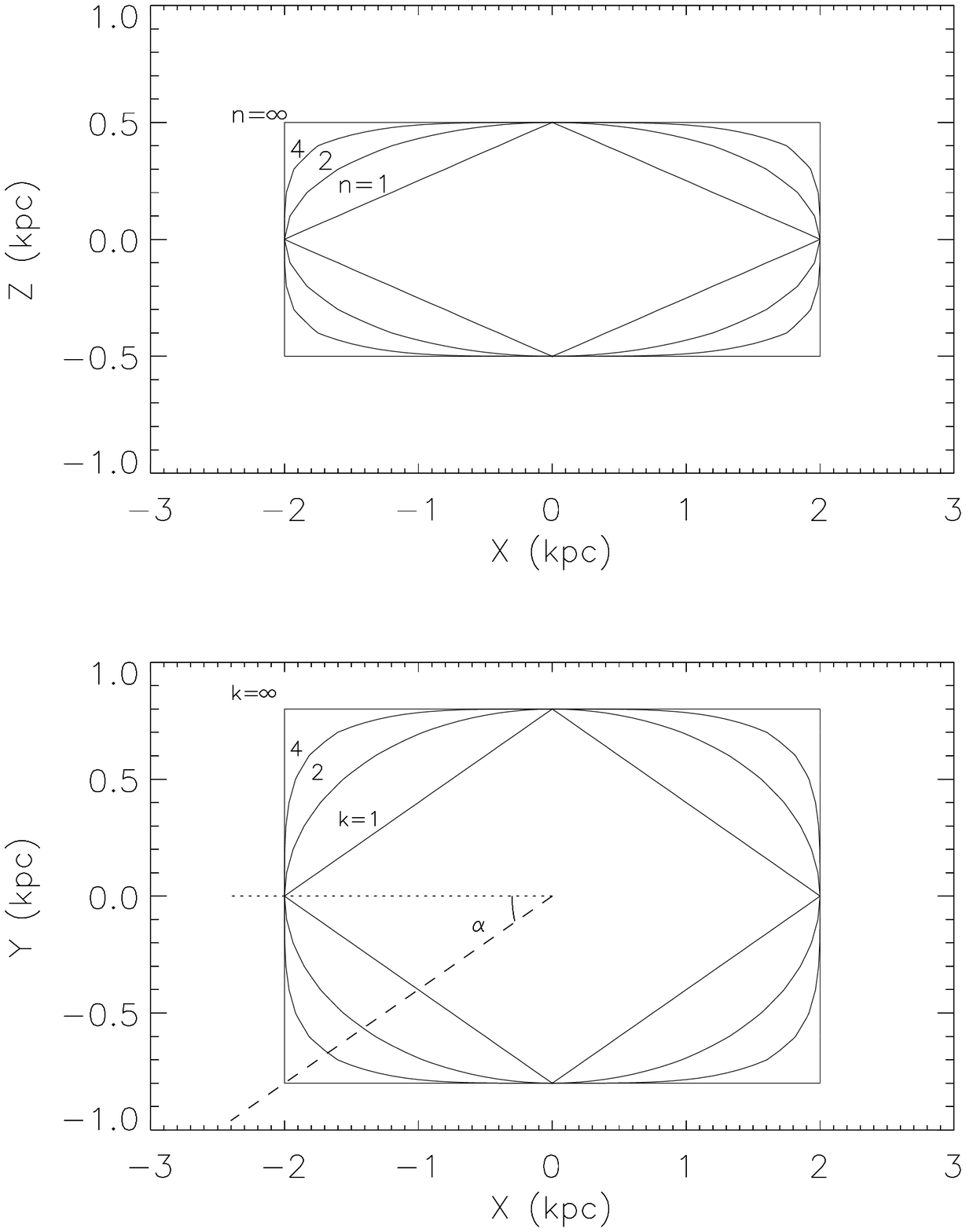}}
\caption{
(a) Some examples of geometrical shapes are shown for the family
of bar models as described by eqs. (\protect\ref{rhoxyz}) and
(\protect\ref{lambdan}). For illustration we used a bar with axis scale
lengths $x_0:y_0:z_0=2:0.8:0.5$ (kpc), similar to the scale lengths of
the COBE constrained GE model (cf. Table 1.) The shapes in the X-Y
plane are determined by $k$ while the shapes in the X--Z (also 
Y--Z but not shown) plane are
determined by $n$. The shape of the bar evolves from being diamond,
ellipsoidal, boxy to rectangular when $k$ or $n$ are increased in the
sequence of 1, 2, 4, and $\infty$. The dashed line indicates
our line of sight to the center, which makes an angle $\alpha$
with the major (X) axis of the bar.  
Observationally, $\alpha \sim
10^\circ-45^\circ$ with the near end of the bar in the first Galactic
quadrant. }
\epsfysize=10cm
\centerline{\epsfbox{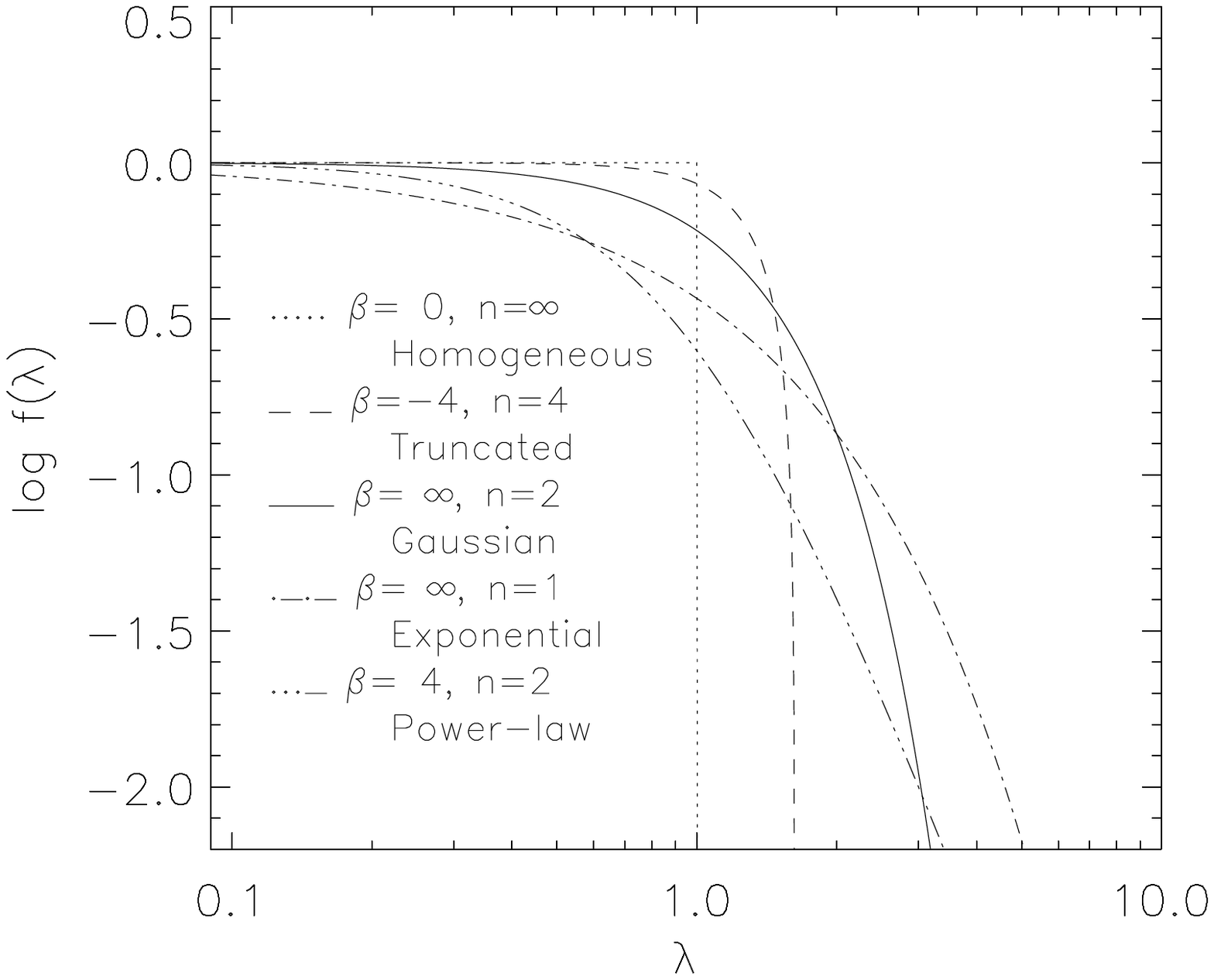}}
(b) Some examples of radial profiles are shown for the same
family of bar models. The density profiles are described by two
parameters $n$ and $\beta$ in the radial coordinates expressed by
$\lambda$ (cf. eqs. \protect\ref{rhoxyz} and \protect\ref{lambdan}). All the density
profiles are normalized to the central density.
\end{figure}

Similar to the Gaussian ellipsoidal bar model, one can show that for this
family the optical depth at any minor axis field depends on $z$ as:
\beq\label{taupower}
\left< \tau(0,z) \right> =  {4 \pi G \over c^2}  {M \over z_0}~
\xi_0(\beta,n) ~ H(k,{y_0 \over x_0},\alpha)
\left[1+ {1 \over \beta-3}
\left({|z| \over z_0}\right)^n\right]^{-{\beta -2 \over n}},
\eeq
where
\beq\label{H}
H(k,{y_0 \over x_0},\alpha)
={ 1 \over K(k)}
\left[\left({y_0 \over x_0} \cos^2\alpha\right)^{k \over 2} + 
\left({x_0 \over y_0} \sin^2\alpha\right)^{k \over 2} \right]^{-k/2}
={ 1 \over K(k)}~
{ 2 \left({u^{k \over 2}+ u^{-{k \over 2}}}\right) ^{-{2 / k}}
 \over \sin 2\alpha },
\eeq
$\xi_0(\beta, n) \approx 
0.6 \pm 0.15 $ is a slow varying function of $\beta$ and $n$, $u$
is as defined in eq. (\ref{IG}), 
$K(k)=4k^{-1} B(k^{-1},k^{-1})$,
and $B$ is the complete Beta-function. For clarity,
we have deferred the somewhat lengthy definition of $\xi_0$ (together with
several useful approximations) until Appendix A.
Eq. (\ref{taupower}) reduces to
eq.~(\ref{taugauss}) for a Gaussian ellipsoidal bar
when $(\beta,n,k) =(\infty,2,2)$,
with $\xi_0 = \pi^{-1/2} , H(2, y_0/x_0, \alpha)= I(y_0/x_0, \alpha)$.

For ellipsoidal models with 
$n=k=2$, an analogous microlensing map as eq. (\ref{gausstaumap}) can
be derived.  We find 
\beq\label{taumap}
\left< \tau(y,z) \right>=\left< \tau(0,0) \right> \left[1+ {1 \over \beta-3}
\left({y^2 \over {y^\prime_1}^2} 
+{z^2 \over z_0^2} \right)
\right]^{-{\beta -2 \over 2}},~~ \mbox{\RM{ if $n=k=2$, }}
\eeq
where $y^\prime_1$ is given by eqs. (\ref{gausstaumap}) and (\ref{x1y1z1}).
Along the line with a constant $z$, 
the optical depth is the largest on the minor axis $y=0$.

Eq. (\ref{taupower}) is our main analytical result. In the rest
of this section, we will study the dependence of the optical depth on
the parameters.

\subsubsection{Dependence on Radial Density Profile and Latitude}

(1) The optical depth due to the bar lenses, $\tau$, is proportional
to the total bar mass $M$, as expected.

(2) $\tau$ falls off on the minor axis as 
$\left[ {\rho(z) \over \rho(0)} \right]^{1 -{ 2 \over \beta}}$
(cf. eqs. \ref{lambdan} and \ref{taupower}).
When $\beta>3$, the model is a power-law with a finite mass,
$\tau$ falls off slower than the density.
When $\beta \le 0$, the model is truncated at a finite size,
$\tau$ falls off faster than the density. When
$\beta \rightarrow \infty$, $\tau$ falls off as the density.

(3) For a fixed field $z$ kpc above the plane, $\tau$ first increases
and then decreases with the vertical scale length $z_0$, and 
$\tau$ is maximized when
the scale height $z_0=z$.   This is
because increasing the scale height leads to both a shallower density
gradient and a smaller central density.
To obtain the maximum optical depth, we note that 
in eq. (\ref{taupower}) the factor
\beq
{M \over z_0} \left[1+ {1 \over \beta-3}\left({z \over z_0}\right)^n
\right]^{-{\beta -2 \over n}} \le 
{M \over z} 
\left({\beta-2 \over \beta-3}\right)^{-{\beta-2 \over n}}.
\eeq
The equality sign indicates the maximum and is reached when $z_0=z$
irrespective of $\beta$.

\subsubsection{Dependence on Bar Angle and Shape}

Not surprisingly, $\tau$ depends on some combinations of the angle,
axis ratio and the shape of the bar, more specifically:

(4) For fixed axis ratio and shape, $\tau$ is largest when the bar
points toward us, i.e., when $\alpha=0$.  In this case, $\tau \propto
H(k, {y_0 \over x_0},0) \propto {x_0 \over y_0}$
(cf. eqs.~\ref{taupower} and~\ref{H}).

(5) For a fixed angle $\alpha<45^\circ$, gradually squashing an oblate bulge to
make it increasingly elliptical in the $x-y$ plane leads to a
monotonic increase in the optical depth until
${y_0 \over x_0} \le \tan(\alpha)$.  For $\alpha >45^\circ$,
the optical depth always decreases when $y_0/x_0$ changes from one to zero.
$\tau$ has a maximum at $u=1$
(cf. eq.~\ref{H}),i.e., at ${y_0 \over x_0}=\tan \alpha$.  This is
graphically shown in the upper panel of Fig. 2 for $k=2$ models.
The turn-over for very small axis ratio can be intuitively understood
as a needle-shaped bar pointing sideways has 
zero depth to place the lens and the source 
along the line of sight, hence zero optical depth.
Mathematically, $\tau$ is proportional to
$H$ (cf. eq.~\ref{taupower}), and 
\beq
H(k,{y_0 \over x_0},\alpha)
={ 2 \over K(k)}~
{ \left({u^{k \over 2}+ u^{-{k \over 2}}}\right) ^{-{2 / k}}
 \over \sin 2\alpha } 
\le { 2 \over K(k)}~{ 2^{-{2 \over k}} \over \sin 2\alpha }.
\eeq
So $\tau$ and $H$ are maximized at $u=1$ irrespective of $k$.  

(6) The maximum gain in optical depth of an elliptical
$(k=2)$ bar vs an oblate bulge is given by 
\beq {\tau_{bar} \over
\tau_{bulge} } \le { 1 \over \sin 2 \alpha},
\eeq
where the maximum is at ${y_0 \over x_0}=\tan\alpha$.
For example a bar with $\alpha=\arctan^{-1}{y_0 \over x_0}=15^\circ$
gains by a factor of 2.  But for $\alpha \ge 45^\circ$ the oblate model
predicts higher optical depth than bar models.

(7) Other things being equal, $\tau$ is a slow varying function of
$n$.  The ratios for $n=1,2,4$ (diamond, ellipsoidal, boxy) for $\beta
\rightarrow +\infty$ are about $1.3:1:0.95$, while for $\beta=4$ the
ratios are $1.68:1:0.82$.

(8) Similarly, for fixed axis ratio and angle, varying $k$, $\tau$ changes by
less than a factor of 2 (see Fig. 2, lower panel). The ratios
of the optical depth $\tau$ for the diamond shape $(k=1)$ to that
for an ellipsoidal shape $(k=2)$ are $1.57$, $1$, and $0.79$ at
$u={y_0/x_0 \over \tan(\alpha)}=0$, $0.31$, and 1 respectively.  
The variation is the same
for $u>1$ due to the symmetry between $u$ and $u^{-1}$ (cf. eq.
[\ref{taupower}]).

(9) Of the whole class of models with different central densities and
shapes, the optical depth is the largest for a homogeneous rectangular
(HR) bar.  The maximum optical depth is reached when
$\alpha=\tan^{-1}\left({y_0 \over x_0}\right)$, i.e., when the line of
sight coincides with the diagonal line of the homogeneous rectangular
bar.  The microlensing map of this model is given by equations in
Appendix B.  A comparison of the homogeneous rectangular (HR)
model with models of other shapes is in \S~\ref{optimalsection}.

\begin{figure}
\epsfysize=15cm
\epsfbox{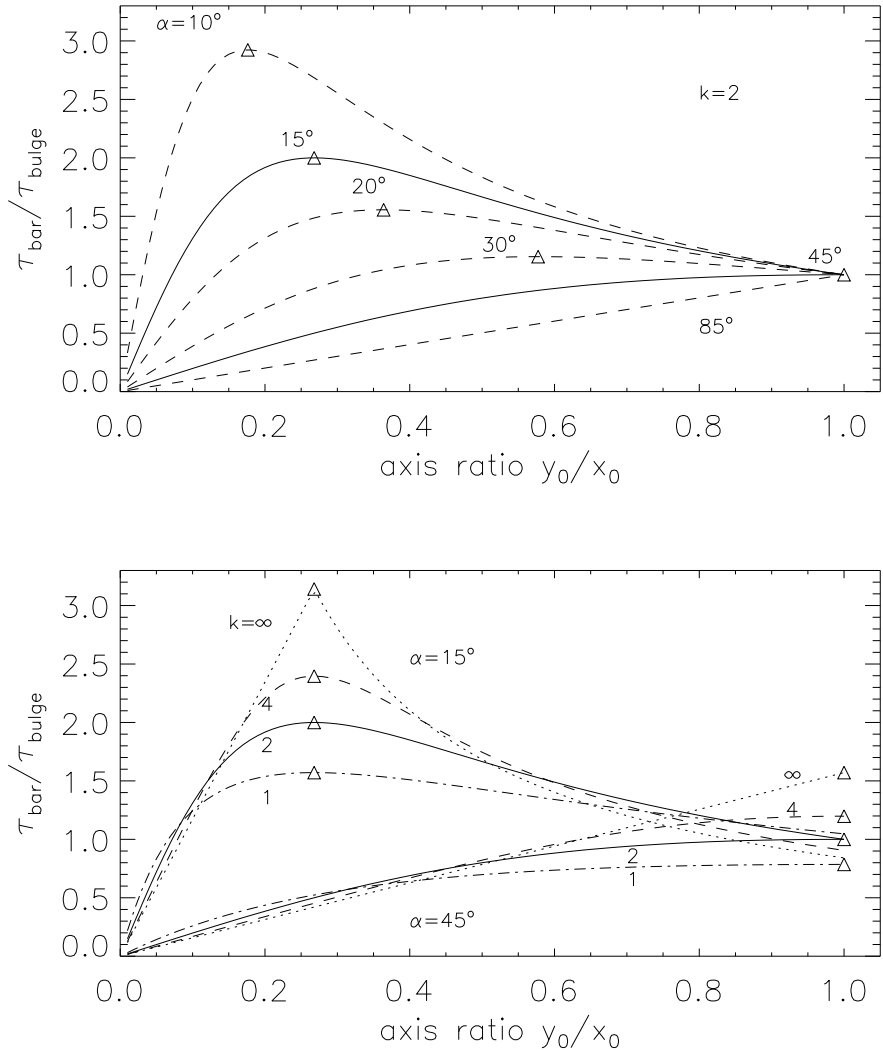}
\caption{
shows the ratio of optical depths of a bar 
vs an oblate bulge as
a function of the bar angle and shape.  
The upper panel shows that for
an elliptical bar
($k=2$), the optical depth peaks at an axis ratio ${y_0 \over x_0} =
\tan(\alpha)$ if the bar angle $\alpha \le 45^\circ$.  
For $\alpha \ge 45^\circ$, the optical depths increases monotonically
with the axis ratio $y_0/x_0$, and is always smaller than that of an
oblate bulge model.  Note
an elliptical bar becomes an oblate bulge at ${y_0 \over x_0} = 1$.
For a bar angle $\alpha=15^\circ$, the lower
panel shows that an elliptical bar gives a larger peak (maximum)
optical depth than that for a 
diamond shaped bar ($k=1$), but less than those for a boxy bar ($k=4$),
and a homogeneous bar. The
positions of the maximum optical depths are indicated by
triangles. They always occur at the axis ratio ${y_0 \over x_0}=\tan(\alpha)
=0.27$ for $\alpha=15^\circ$.
}
\end{figure}

\section{Application to the Galactic Bar}

The formalism we have developed can be applied to any edge-on bar.
In this section, we will concentrate on the Galactic bar, where most
microlensing events are observed (Alcock 1995b; Udalski et al. 1994;
Alard et al. 1995).  As the optical depth is a function of the bar
parameters such as the total mass, bar angle and axial scale lengths,
we therefore first outline the uncertainties in these parameters
before we proceed to estimate the optical depth.

\subsection{Uncertainties in the Bar Parameters}

Dwek et al. (1995) carried out a systematic study of the bar
parameters by fitting the COBE map. A few models can fit the data
well, i.e., the parameters are not well constrained. For
non-axisymmetric models, the bar angle lies $10^\circ \la \alpha \la
45^\circ$, and the axis ratio $y_0/x_0= 0.15-0.5$, and the vertical
scale height $z_0=200-500$ pc. In Table I, we have listed four models
from Dwek et al. (1995) that can be described by our parametrization
in eqs. (\ref{rhoxyz}) and (\ref{lambdan}).
Except for the power-law exponential model, 
the errorbars for the parameters are quite small (cf. Table 1), while
the variations from model to model are quite large. For each model,
we will take the best-fit (central) parameter values in the table and evaluate
the optical depth. We do not attempt to do a full error analysis due
to the correlation of errors between parameters.
Out of the four models, the Gaussian ellipsoidal model
provides the best fit to the COBE map, while the exponential diamond (ED),
oblate bulge (OB) and power-law ellipsoidal (PE) profiles become progressively
less satisfactory. Note that the oblate bulge model is an
axisymmetric model.
Two additional models are listed in Table 1, one being a truncated boxy (TB)
bar model, and the last one a homogeneous rectangular (HR) bar.
These models are interesting since extragalactic bars often have a flat
density profile and a sharp truncation near corotation 
(Sellwood \& Wilkinson 1993).  For these
two models, the bar parameters are not available and we only consider
their optimal lensing configurations. These six models
should offer a broad survey of the uncertainty in the optical depth.

\begin{table}
\caption{Parameters for Six Galactic Bar Models}
\begin{tabular}{lllllll}
Model & OB (G0) & GE (G1)	& ED (E1)	& PE (P3)& TB & HR	 \\
Description	&  	& Gaussian, 	& Exponential,	& Power-Law,	& Truncated,	& Homogeneous,	\\
	& Oblate 	& Ellipsoidal	& Diamond 	& Ellipsoidal
		& Boxy		& Rectangular	\\
$(\beta, n, k)$ & $(\infty, 2, 2)$ & $(\infty, 2, 2)$ & $(\infty, 1, 1)$ &
$(4, 2, 2)$	& $(-4, 4, 2)$ & $(0, \infty, \infty)$	\\
$\alpha(^\circ)$	&---	& $11 \pm 2.4$	& $24.4 \pm 6.7$	&
$45.4 \pm 33.3$	& ---	& ---	\\
$x_0$ (kpc)	& $0.91 \pm 0.01$	& $2.08 \pm 0.06$	& $1.64
\pm 0.06$	& $0.90 \pm 0.2$	& ---	& ---	\\
$y_0$ (kpc)	& $0.91 \pm 0.01$	& $0.75 \pm 0.01$	& $0.61
\pm 0.11$	& $0.23 \pm 0.98$	& ---	& ---	\\
$z_0$ (kpc)	& $0.51 \pm 0.01$	& $0.45 \pm 0.02$	& $0.31
\pm 0.01$	& $0.28 \pm 0.04$	& ---	& ---	\\
$\xi_0(\beta,n)$& 0.564	& 0.564	& 0.75	& 0.477	& 0.554	& 0.667	\\
$\xi_4(\beta,n)$& 1.772 & 1.772	& 2.55	& 2.094	& 1.376	& 1	\\
$H(k,{y_0 \over x_0}, \alpha)$	& 0.159	& 0.354	& $0.165$	&
0.08 & --- & ---	\\
$\langle\tau_{-6}(0,0)\rangle$	& 2.9	& 7.3	& 6.5	&
$2.2$	& ---	& ---	\\
$\langle\tau_{-6}(0,-2^\circ)\rangle$	& 2.4	& 5.8	& 2.5 	&
$1.1$		& ---	& ---	\\
$\langle\tau_{-6}(0,-4^\circ)\rangle$	& 1.5	& 3.1	& 0.94 	&
$0.41$		& ---	& ---	\\
$\langle\tau_{-6}(0,-6^\circ)\rangle$	& 0.64	& 1.1	& 0.38 	&
$0.20$		& ---	& ---	\\
\end{tabular}

\medskip
Six models for a very massive ($M=2.8\times 10^{10} M_\odot$) 
Galactic bar. The radial profiles
and geometrical shapes of the models are described in the second
and third rows. The initials of these descriptions in the first row 
are used to label
the curves in the figures. The first four models are the same as
the G0, G1, E1, and P3 models in Dwek et al. (1995).
$(\beta, n, k)$ are parameters that
describe the shape of the bar (cf. eq. \ref{lambdan}).
$\alpha$ is the angle between the direction to the Galactic centre and
the major ($X$) axis of the bar.  $x_0, y_0, z_0$
are the scale lengths of the three principal axes of the bar.
The parameters for the first four models are taken from Table 1
in Dwek et al. (1995) at 2.2 $\mu{m}$ with $R_{max}=\infty$.
$\xi_0(\beta, n)$ , $\xi_4(\beta, n)$ and $H(k, y_0/x_0, \alpha)$
are functions
of $\beta$, $n$, $k$, $\alpha$ and $y_0/x_0$. These are used in eqs. 
(\ref{taupower}) and (\ref{correct}) to evaluate the optical depth. 
The last four rows list the predicted 
optical depths due to lenses in the bar at four fields; the values
would scale linearly for other choice of 
the bar mass $M$ (eq. \ref{mbar} gives the
range of $M$).
\end{table}

The mass of the Galactic bar cannot be derived directly from the
COBE map. However, it can be obtained by fitting either the gas
rotation curve or the stellar velocity dispersions.
Generally axisymmetric models such as the Kent's (1992) oblate
rotator model (cf. the oblate  model in Table 1) fail to 
match the gas rotation curve either at $R \approx 0.5$ kpc or at $R
\approx 3$ kpc. Triaxial models
with and without pattern rotation can be made consistent with the gas
rotation curve. A rough limit on the mass of the bar can be set
as follows:

If the bar is contained within a corotation radius of 
$R_{\mathrm{cor}} \approx 2.4$ kpc
(Binney et al. 1991), then the enclosed dynamical mass, $M_d + M$, satisfies
\beq
\eta {G (M_d + M ) \over R_{\mathrm{cor}} }  =  V^2_c(R_{\mathrm{cor}}),
\eeq
where $M_d$ is the mass contributed by the disk within $R_{\mathrm{cor}}$, and
$V_c(R_{\mathrm{cor}}) \approx 195$ km s$^{-1}$ is the circular velocity at the
corotation radius. The parameter $\eta$ is a factor that describes
the enhanced gravity of a flattened or barred system relative
to a spherical system. Since $\eta \ge 1$ and $M_d \sim (10\%-20\%) M >0$,
we derive a loose upper bound for the bar mass,
\beq
M \la 2.2 \times 10^{10} M_\odot
\eeq
by setting $\eta=1$ and $M_d=0$.

More detailed gas modelling that matches the non-self-crossing
closed orbits with the gas terminal velocity gives a very similar value.
Assuming different pattern speeds, viewing angles, and bar
density profiles, several authors find $M=(2.2 \pm 0.2) \times 10^{10}
M_\odot$ (Gerhard \& Vietri 1986; Binney et al. 1991; ZRS96)

The mass obtained by different authors using the stellar velocity
dispersions of bulge stars has a larger scatter. This is
mainly due to different levels of
sophistication in the dynamical models.
The simple oblate rotator model by Kent (1991) predicts a low value,
$M=1.8 \times 10^{10} M_\odot$ based on the pre-COBE data. The detailed
steady state bar model of Zhao (1996), which fits the COBE map with a
positive definite distribution function, gives $M=(2.2 \pm 0.2) \times
10^{10} M_\odot$.  Straightforward but less certain application of
the Virial theorem predicts between $M=1.8 \times 10^{10} M_\odot$ with
no pattern rotation of the bar (Han \& Gould 1995), and
$M=2.8 \times 10^{10} M_\odot$ with a big pattern rotation (Blum 1995).

To summarize, the predicted mass of the bar by different authors
is in the range of
\beq \label{mbar}
1.8\times 10^{10} M_\odot \le M \le 2.8 \times 10^{10} M_\odot.
\eeq

\subsection{Optical Depth of Disk Lenses}

To compare our bar models with observations, 
the contribution from the disk lenses must be added.
It is well known that a standard disk model predicts $\tau_{\mathrm{disk},
-6}\approx 0.66$ at Baade window, too small to explain the observed
high optical depth. However, our predictions for disk lenses are
still uncertain due to lack of constraints of faint disk sources
at several kpc from the Galactic center. ZSR95 find that the optical
depth of the disk is in the range of $0.37-0.87$ for Baade
window by extrapolating the local disk density 
with a double-exponential disk model 
with different disk scale length, scale height and local density normalization.
The lower end is from a disk with a central hole
as suggested by Paczy\'nski et al (1994b) based on 
color-magnitude diagrams of OGLE stars.
%
%
A much higher disk optical
depth ($\tau_{\mathrm{disk,-6}}=2)$ 
would require the disk to make up most of the rotation curve,
which appears to be in conflict with the lack of faint stars
or brown dwarfs in the solar neighbourhood (Gould, Bahcall, \& Flynn 1996). 
As the effect of the disk is small, in this paper,
we will simply adopt 
\beq
\tau_{\mathrm{disk},-6} \approx 0.35-1,
\eeq 
and completely ignore its $z$ dependence.
This range is indicated by two thick horizontal bars
at the bottom left in Figures 3--6.

\subsection{Optical Depth of the COBE Bar: Red Clump Sources}

Having discussed the possible ranges of the bar parameters and 
the disk contribution, we
can now investigate the optical depth of the COBE bar. 
For bright red clump giants in the bulge, we can apply
eq. (\ref{taupower}), which we rewrite as
\beq \label{tau0}
\langle\tau_{-6}(0,z)\rangle = 2.1~{M \over 2 \times 10^{10} M_\odot}
{0.5 {\mathrm{kpc}} \over z_0}
{ \xi_0(\beta,n,k) \over 0.564}
{H(k,{y_0 \over x_0},\alpha) \over 0.159}
\left[1+ {1 \over \beta-3}\left({|z| \over z_0}\right)^n\right]
^{-{\beta -2 \over n}},
\eeq
where we have normalized the quantities $\xi_0(\beta,n,k)$ and $H(k,y_0/x_0,
\alpha)$ to the values for an oblate bulge, i.e., 
for $y_0/x_0=1$ and $n=k=2$.  

Table 1 gives the optical depths for four latitudes along the
minor axis for the four models taken from Dwek et al. (1995). The field
at $(0^\circ, -4^\circ)$ is close to Baade window. The values are for
$M=2.8\times 10^{10} M_\odot$, at the upper end of eq.
(\ref{mbar}), but this can be scaled linearly for other $M$. There are two
things worth pointing out: first, at the same field 
the optical depths of different models differ by 
a factor 3--5;
second, the optical depth drops with
latitude for each model.  From the Galactic center to Baade window,
the decrease is about a factor of 6 for the exponential diamond (ED)
and the power-law ellipsoidal (PE) models, but only about
a factor of 2--3 for other models. The fast decrease of the exponential
diamond and power-law ellipsoidal models are obviously due to their
steep density profiles (cf. Fig.1b and Table 1). 

These points are further illustrated in Fig. 3, where we have plotted
the variation of the optical depth of the bar along the minor
axis. The black dots with $2\sigma$ errorbars are the inferred optical
depths for the low latitude and high latitude clump giants (Alcock et
al. 1995b). Notice the data is suggestive of a decrease in the optical
depth with latitude.

\begin{figure}
\epsfysize=15cm
\epsfbox{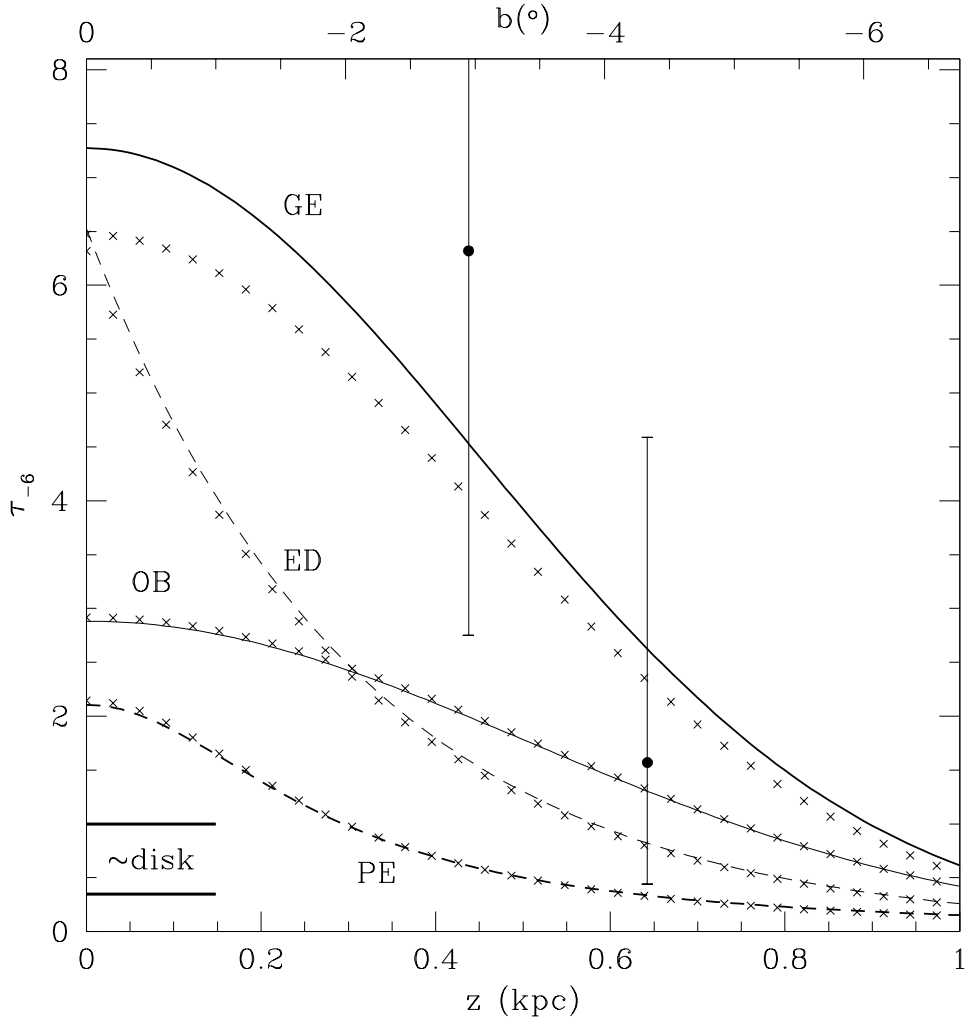}
\caption{
Optical depths on the $z$ (minor) axis are shown for four
COBE constrained bar models (columns 2 to 5 in Table 1)
for clump sources. The $z$ coordinate and latitude are shown at the bottom
and top axes respectively. The abbreviations for the models
are given in Table 1.
All the optical depths are for a very massive bar with $M=2.8 \times
10^{10}M_\odot$, but can be scaled linearly for other masses (the
range is given in eq.~\protect\ref{mbar}). The approximate range of
the optical depth contributed by the disk are indicated with two
short thick lines at the bottom left. The total optical can be calculated
by first linearly scaling the plotted values for different $M$ and then add
the disk contribution. The two black dots are the optical depths
observed for the low and high latitude clump giants (Alcock et al. 1995b).
The $2\sigma$ errorbars are also plotted. The crosses indicate full numerical
calculations for each model.
}
\end{figure}

Now let us examine Fig. 3 more closely. We notice that the
errorbars are very large, which is hardly surprising because the two
subsamples have only 10 and 3 events, respectively! Still, out of the
four models, only the Gaussian ellipsoidal (GE)
model can produce an optical depth as high as
$\tau_{-6}=6$. This is achieved only when the bar is very massive,
with $M = 2.8\times 10^{10} M_\odot$ and with a large disk contribution,
$\tau_{\mathrm{disk},-6}=1$. For the same
condition, the oblate bulge and exponential diamond
models are only at the $2\sigma$ lower limit.
The power-law ellipsoidal model produces an optical depth
simply too small to be consistent with observation. This is
mainly due to the large bar angle, $\alpha \approx 45^\circ$, 
rendering a very inefficient lensing configuration.

\subsection{Optical Depth of the COBE Bar: Main Sequence Sources}

We can carry out a similar analysis for the main sequence stars.  The
analytical formulae that apply to these stars are given in
eqs. (\ref{correct}) and (\ref{xi4}). But in Fig. 4, we only show
the numerical results from eq. (\ref{tauav}). Each shaded region
corresponds to one model, with the upper and lower curves for
$\beta_s=1$ and $\beta_s=2$ respectively.

Quite strikingly the optical depth decreases with increasing
$\beta_s$, which may explain the smaller optical depth for main
sequence stars $\beta_s>0$ in the MACHO sample than for the clump
giants $\beta_s=0$.
%
%
For the Gaussian ellipsoidal model, the optical depth drops by a
factor of 2 from $\beta_s=0$ to $\beta_s=2$, while the power-law
ellipsoidal model is almost unchanged. For the other two models, the
decrease is about (20-50)\%. Increasing $\beta_s$ effectively moves
the average source closer to us and reduces the optical depth by a
factor proportional to ${x_1 \over R_0}\beta_s$, where $x_1$ is the scale
length in the line of sight (cf. eq.\ref{x1y1z1}). 
The Gaussian ellipsoidal model has the
largest major axis length and smallest angle $\alpha$, hence the
largest $x_1$, the reduction for this model is therefore the strongest. 
The decrease of $\tau$ for intrinsically faint sources
can potentially provide interesting limits on the source luminosity function
if the bar is sufficiently extended.

\begin{figure}
\epsfysize=15cm
\epsfbox{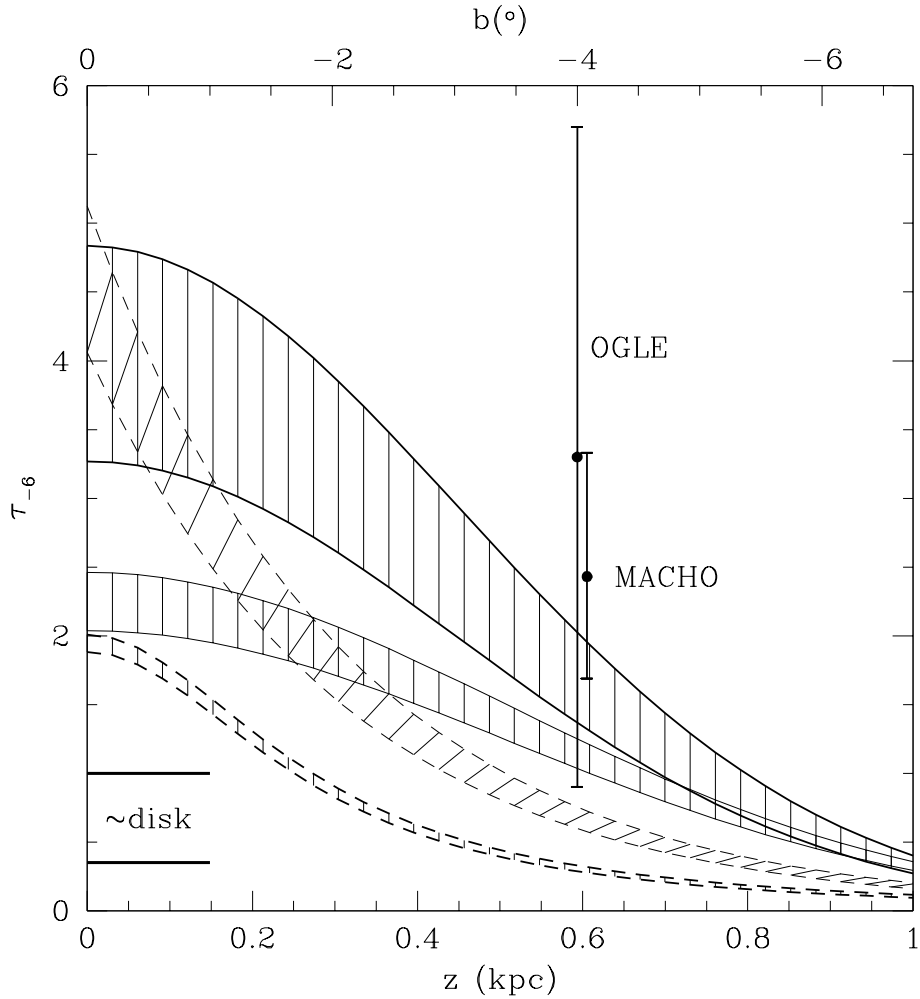}
\caption{Similar to Fig. 3, but for minor axis
optical depth of four 
COBE models with source being bulge main sequence sources. 
Each shaded 
region corresponds to one model, with the top curve for $\beta_s=1$
and the lower curve for $\beta_s=2$.
Again, a very massive bar with $M=2.8 \times 10^{10}M_\odot$ is used.
The optical depths for the full MACHO and OGLE samples are shown
together with their $2\sigma$ errorbars.
}
\end{figure}

%

In the same figure, the optical depths from 41 MACHO events and 9 OGLE
events are shown with their $2\sigma$ errorbars.  With a bar mass of
$M=2.8\times 10^{10} M_\odot$ and a disk contribution of $10^{-6}$,
only the Gaussian ellipsoidal and exponential diamond models are
within the $2\sigma$ errorbars of the MACHO sample. For the diamond
exponential and power-law ellipsoidal models, the optical depths are
too small even if we adopt a large bar mass and substantial disk
contribution.

\subsection{Optimal Lensing Configuration of the Bar}\label{optimalsection}

We have so far studied the COBE constrained bar models. Three out of the
four models have difficulties in reproducing the observed high optimal
depth. It is therefore interesting to examine whether these bar models --
disregarding the COBE constraints-- can produce the high optical depth
at all. In other words, we want to know the maximum optical depth for
each model by adopting the optical lensing parameters. For
this exercise, we will concentrate on the clump giants as these are
less prone to systematic effects such as blending and also because they
appear to have larger optical depths. As the observed fields of MACHO
and OGLE are centred near Baade window, we choose to maximize the
optical depth at this window for each model.  

For each of the six sets of $(\beta, n, k)$ in Table 1, the optical
depth at Baade window depends linearly on the total mass and
on some combinations of the angle $\alpha$, the axis
ratio $y_0/x_0$, and the vertical scale height $z_0$.  For the Baade
window, the most favorable lensing configuration is when the vertical
scale height $z_0=z_{BW} \approx 590$ pc (see the discussion at the
end of \S 2.3.1), and the axis ratio $y_0/x_0= \tan \alpha$. 
 
In Fig. 5, we plot the optical depth for the above six bar models
each at its most favorable configuration ($z_0=z_{BW}$ and $y_0/x_0=
\tan \alpha$) as a function of the bar angle $\alpha$.  Clearly for
each model the optical depth drops quickly when $\alpha$ is increased
from $0^\circ$ to $20^\circ$, after which the curves become nearly
flat. This is simply due to the factor $1/\sin(2\alpha)$ in
$H(k, y_0/x_0, \alpha)$ (cf. eq. \ref{H}).

\begin{figure}
\epsfysize=15cm
\epsfbox{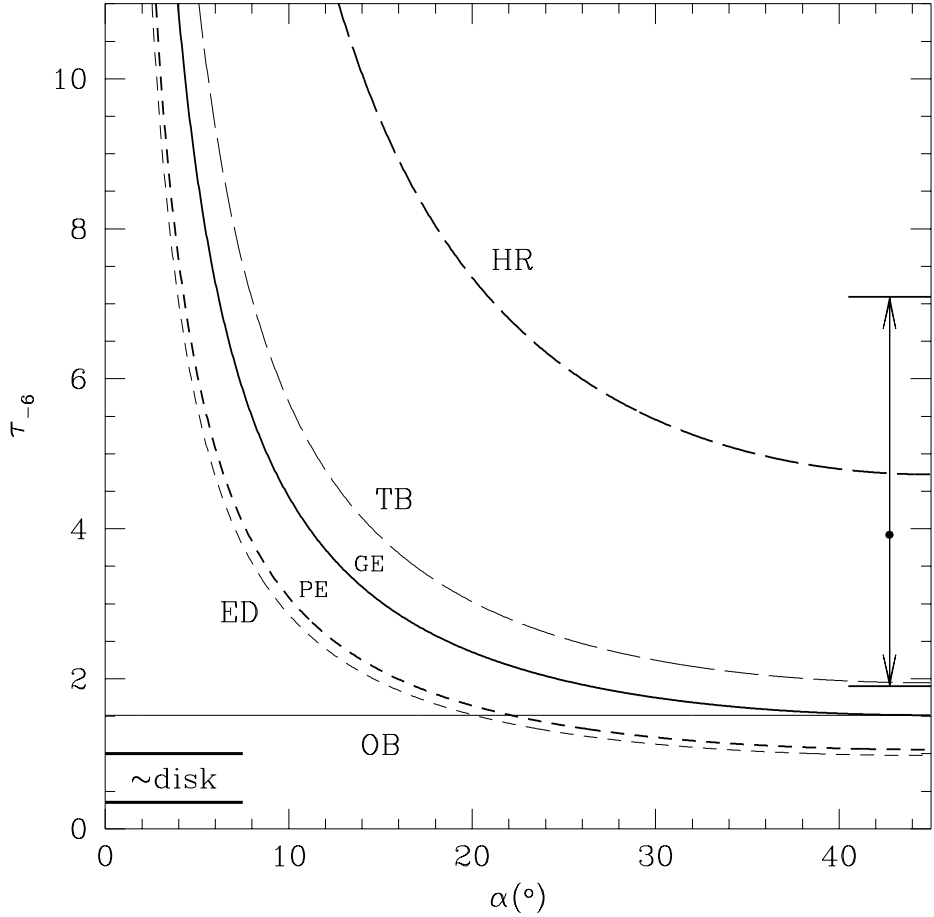}
\caption{
Optimal optical depths at Baade window are
shown as a function of the bar angle $\alpha$, for a very massive bar
$M=2.8\times 10^{10}M_\odot$. To achieve the optimal optical depth,
the vertical scale height is fixd to be $z_0=590$ pc, and the axis ratio
$y_0/x_0=\tan \alpha$. The abbreviations for the models are listed in Table
1. The observed range of optical depth for the clump giants (Alcock et al.
1995b) is shown by two horizontal bars with arrows with the black dot
indicating the central value. The approximate optical depth contributed by the
disk lenses are plotted with two short thick lines at the bottom left. The
total optical depth
can be calculated by first scaling the plotted values linearly
for different $M$ and then adding the disk contribution.
}
\end{figure}

The six models in Table 1 have different radial profiles and boxyness
due to their different $(\beta, n, k)$. Fig. 5 shows that
the boxy bars with a flat density profile have an interesting
signature of in optical depth. For any given
bar angle the optical depth is the largest for the homogeneous
rectangular bar.  Intuitively this can be understood as follows. A
large optical depth to the bulge requires putting as many lenses as
possible at positions as distant as possible from the source.
A model with a flat density profile in the line of sight
places more lenses at the near side and more sources at the far side than a
model with a steep density profile, hence a model with
a flat profile is a more favorable lensing configuration.
In addition, for a homogeneous bar the number of lenses does not
decrease as one moves up on the minor axis before reaching the boundary.
A rectangular bar furthermore stretches the distance between the lens and
source even more as it has the longest contributing line of sight. A diamond
shaped model, on the contrary, has the shortest contributing line of
sight. These points become clear by comparing the length of segments of the
dashed line intercepting the rectangular contour and the diamond contour in
the bottom panel of Fig. 1a. Using the same argument, one can
understand why the truncated boxy model (TB model in Fig. 5) also
has a large optical depth.

This figure also points out the promise of the microlensing technique.
If the present observed mean value of the optical depth is the true
value, we see from Fig. 5 that all the bar models must have angle
within $20^\circ$ except the extreme homogeneous rectangular (HR)
model. Unfortunately, if the lower $2\sigma$ bound of the optical
depth is true, and if one adopts a massive bar and a large disk
contribution ($\tau_{\mathrm{disk},-6}=1$), then the current sample provides
little constraint on the bar angle.
The situation could be improved significantly with a factor of four
increase in the sample size; the errorbar will be reduced by a
factor of two. If the optical depth $\tau_{-6} \ge 3.9 $ were
confirmed, then the most plausible models would be a $M \ge 2\times
10^{10}M_\odot$ truncated boxy (TB) bar with an angle
$\alpha=10^\circ-20^\circ$, and axis ratio ${z_0 \over x_0} \le {y_0 \over
x_0} \approx \tan(\alpha)$ plus a modest contribution of disk lenses.
The optical depth $\tau_{-6} \sim 12$ at the high $2\sigma$ end of the
MACHO clump giant sample is likely unphysical since it would argue for
one of the following: (a) an extremely elongated bar with ${y_o \over
x_0} =\tan(\alpha)\approx 0.05$ and the line of sight almost coincides
with the major axis of the bar: $\alpha=2^\circ-5^\circ$; (b) an very
elongated homogeneous rectangular bar with $\alpha \approx 10^\circ$; (c)
a drastically different picture of the mass distribution in the
Galaxy. The first two possibilities are probably ruled out by the COBE
map, while the third one will have a host of difficulties with the gas
and stellar kinematics.

In Fig. 5, we have taken $\alpha$ as a complete free parameter, and
the axis ratio is set such as to achieve the maximum optical
depth. Now we want to examine possible constraint for the axis
ratio. To do this, we take $\alpha=10^\circ$, i.e., at the lower end
of the inferred bar angle (cf. Table 1), keeping $z=z_0$, and study
the variation of optical depth with the axis ratio. The results are
plotted in Fig. 6.  Big optical depth would imply that the axis
ratio ${y_0 \over x_0}$ is close to the optimal value $\tan(\alpha)$.
Unfortunately the current data gives only a rather weak limit at 95\%
confidence level.  However, if we take the mean value of the MACHO
experiment and a disk contribution of $10^{-6}$, we can place
non-trivial constraints on the axis ratio.  For example, for the GE
model, we will limit the axis ratio to be $0.05 <y_0/x_0 < 0.5$; the
observed value 0.35 for GE model 
(cf. Table 1) is within this limit. More definite
limits are only possible when the sample size is increased a few
times and the errorbars are reduced.

\begin{figure}
\epsfysize=15cm
\epsfbox{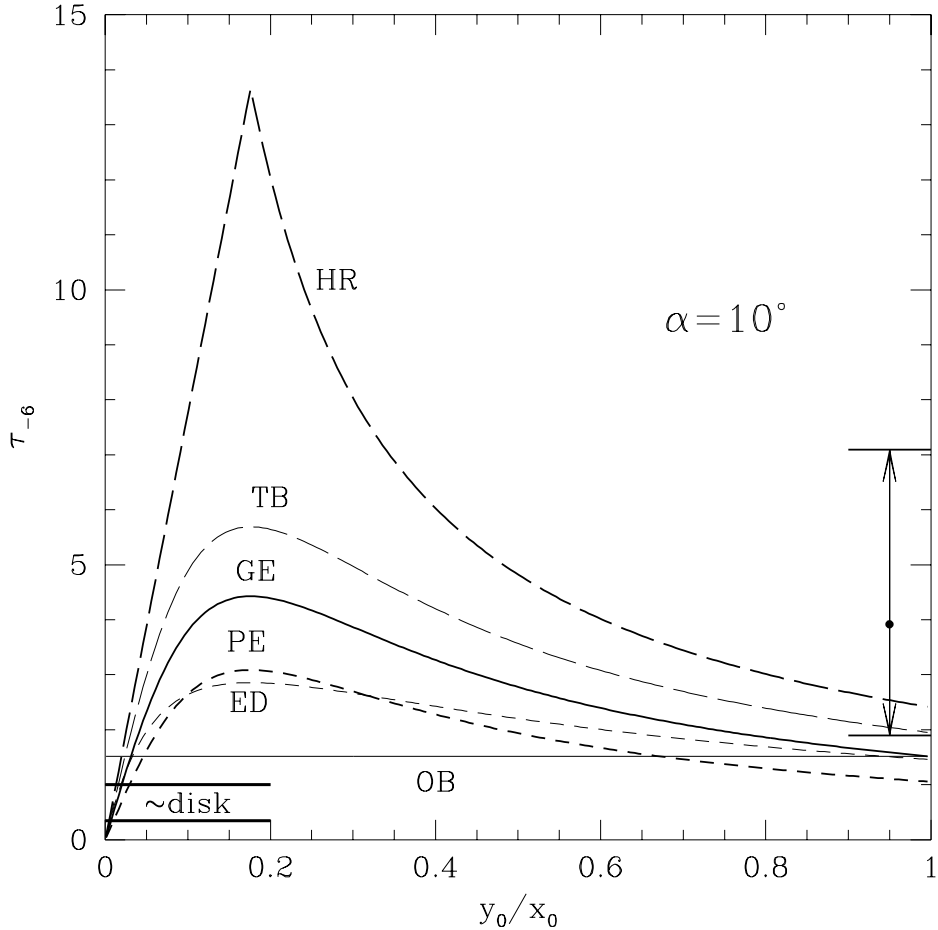}
\caption{
Optical depths at Baade window are shown as a function of
the axis ratio ${y_0 \over x_0}$ 
for a very massive bar, $M=2.8\times 10^{10}M_\odot$,
 with a fixed angle $\alpha=10^\circ$. The vertical
scale height for each model is set to be $z_0=590$ pc.
The abbreviations for the models are listed in Table 1.
$M=2.8\times 10^{10}M_\odot$.
The observed optical depth for all the clump giants by Alcock et al. (1995b)
is shown by two horizontal bars with arrows with the central value
indicated by a black dot.  
The approximate range of the optical depth contributed by the
disk lenses are indicated by two short thick lines at the bottom left. The
total optical can be calculated by first scaling the plotted optical
depth linearly for different $M$ and then adding the disk contribution.
}
\end{figure}

\section{Discussion}

We have evaluated the optical depths for a family of bar models. The
analytical formulae developed are applicable to both the bulge clump
giants and the main sequence stars. We have studied the dependence of
the optical depth on the bar parameters. With about 50 microlensing
events, we have demonstrated that microlensing already provides
interesting constraints on the bar models. For example, the COBE
constrained power-law ellipsoidal (PE) model produces an optical depth just
too small to be consistent with the observed optical depths, primarily
due to the large bar angle, $\alpha \approx 45^\circ$. The
axisymmetric oblate bulge model is also only marginally compatible
with the observation. The Gaussian ellipsoidal (GE) model clearly best
reproduces the observation. It can produce $\tau_{-6}=6$ if the
Galactic bar is very massive ($M=2.8\times 10^{10} M_\odot$) and the
disk contribution is $\tau_{\mathrm{disk},-6}=1$. We believe that the
truncated boxy (TB) model is also a viable model. It is therefore desirable
to constrain the parameters for this kind of finite truncated bar
using the COBE map. From our comparison of the optimal optical depths
with the observations, we found the high optical depth already
suggests a less than $20^\circ$ angle between the long axis of the bar and
our line of sight to the Galactic centre, and that the axis ratio of
the bar is close to the optimal (cf. Fig. 5 and 6).

In our comparison between the prediction and the observation, we have
neglected a central nucleus and the dependence of $\tau$ on the longitude.
The longitude dependence in general reduces the optical depth. But
as the mean longitude of the MACHO and OGLE fields are quite small, $l
\le 2.5^\circ$, which corresponds to approximately 375 pc, the reduction is
about 20\% for the power-law ellipsoidal (PE) model and 
only about 10\% for all the other models. In our Galactic bar models,
%
%
the density profiles are flat near the Galactic centre,
while a nucleus has long been observed in the infrared (Becklin \&
Neugebauer 1968). But since the MACHO and OGLE fields are all offset
by more than $2^\circ$ from the centre, a nucleus within
$2^\circ-3^\circ$ of the Galactic centre does not affect the
microlensing prediction directly except that the nucleus contributes
to the dynamical mass within 3 kpc.  Including a nucleus would reduce
the predicted optical depth outside the nucleus.  But we estimate that
the mass of the nucleus inside $2^\circ$ is only about 5\% of the
total bar mass, much smaller than the uncertainty in the bar and disk
mass.  So the nucleus can be neglected for predicting optical depth at
any field a few degrees ($\ga 3^\circ$) away from the Galactic
centre.

In this study, we have used the MACHO and OGLE samples
(Alcock et al. 1995b; Udalski et al. 1994). The number of analysed
events (50) is still very limited, therefore the errorbars of the
inferred optical depths are still quite large. The subsample of
(13) clump giant events 
that we used has even larger errorbars. In addition, some of the
long events in the MACHO sample are not well understood and these
contribute about 1/3 of the optical depth (Han \& Gould 1996). Therefore
the current limits on the optical depth have to be interpreted with
some caution. Fortunately, the number of events detected toward the
Galactic bulge is
increasing rapidly with time. In two years, the number of (analysed)
microlensing events is likely to increase by four-fold.
It will then become feasible to analyse the whole microlensing map
of a large fraction of the bulge.  If on the order of
one hundred events are obtained at two or three low extinction
fields close to the minor
axis, e.g., the Sgr I field $(1.4^\circ, -2.6^\circ)$ and Baade window
$(1^\circ, -3.9^\circ)$, then we can measure 
the gradient for the optical depth,
which we can use to distinguish bar models with different
minor axis profile and boxyness (cf. Fig. 3).  
If the optical depth remains high,
then tighter limits on the bar parameters can be derived (cf.
Figs. 5 and 6). By combining the microlensing
map with the event duration distributions, it seems possible to
disentangle the relative contributions of the disk and the bar.
Furthermore, a comprehensive study combining the information in the gas,
stellar kinematics, COBE map and microlensing surveys should
provide precise values on all the bar parameters. We conclude
that microlensing has become a promising and unique tool in probing
the Galactic structures.

This project is partly supported by
the ``Sonderforschungsbereich 375-95 f\"ur Astro-Teilchenphysik'' der
Deutschen Forschungsgemeinschaft. We thank Peter Schneider for comments
on the paper.


\appendix
\section{More Details of the Analytical Models}

In this appendix, we give more technical details for the general family
discussed in \S 2.3.

For the density profile described by eqs. (\ref{rhoxyz}) and
(\ref{lambdan}), one can show that
the central density is given by
\beq
\rho_0 = {M \over x_0 y_0 z_0} {1 \over J(n) K(k) \xi_3(\beta,n)},
\eeq
where
\bey
\xi_3(\beta,n)={|\beta-3|^{3 \over n} \over n} & B({3 \over n},& {|\beta| -3 \over n}) \mbox{\RM{ if $\beta > 3$ }}, \nonumber \\
{|\beta-3|^{3 \over n} \over n}
& B({3 \over n},& {|\beta|\over n}+1) \mbox{\RM{ if $\beta \le 0$ }},
\eey
\beq
K(k) = {4 \over k} B({1 \over k}, {1 \over k}) ~~ \approx {8 k^2 \over k^2+1},
\eeq
\beq
J(n) =  {2 \over n} B({1 \over n}, {2 \over n}) ~~ \approx
{3 n^2 \over n^2 + 2},
\eeq
where $B$ is the complete Beta-function, and 
two approximations are given for $K(k)$ and $J(n)$ that are accurate within
3\% for $k>1$ and $n>1$. 

The optical depth at any minor axis field depends on $z$ as follows:
\beq
{\langle \tau(0,z) \rangle \over \langle \tau(0,0) \rangle }
={\tau(0, 0, z) \over \tau(0, 0, 0)}
 = \left[{\rho(0,0,z) \over \rho(0,0,0)}\right]^{ 1 - {2 \over \beta}} \\
 = \left[1+ {1 \over \beta-3}\left({z \over z_0}\right)^n \right]
 ^{-{\beta -2 \over n}}.
\eeq
The optical depth toward the centre $\left< \tau(0,0) \right>$ and 
the optical depth for a source at the centre $\tau(0,0,0)$ are given by
\beq
\langle\tau(0,0)\rangle = \tau(0,0,0) ~ \xi_1(\beta,n) = ({4 \pi G \over c^2}  {M \over z_0}) \xi_0(\beta,n) H(k,{y_0 \over x_0},\alpha),
\eeq
where the quantities $\xi_0, \xi_1, H$ are dimensionless functions of the
bar angle, shape and radial profile. They are defined as
\beq	\label{xi1f}
\xi_1(\beta,n) = {\int_{-\infty}^{+\infty}d\lambda  ~
f(\lambda) \int_0^{+\infty} d\lambda_1 f(|\lambda-\lambda_1|)\lambda_1 
 \over \int_{-\infty}^{+\infty}d\lambda  f(\lambda) \times 
\int_0^{+\infty} d\lambda_1 f(\lambda_1)\lambda_1}.
\eeq
\bey	\label{xi1ratio}
{\xi_0(\beta,n) \over \xi_1(\beta,n)} & = & {n \over 2 |\beta-3|^{1 \over n}
 B({1 \over n}, {|\beta| -3 \over n})},
\mbox{\RM{ if $\beta>3$ }}, \nonumber \\
 &= & 
{n \over 2 |\beta-3|^{1 \over n} B({1 \over n}, {|\beta|+2 \over n}+1 )},
\mbox{\RM{ if $\beta \le 0$ }} \\
 & \rightarrow & { n^{1-1/n} \over 2\Gamma(1/n)},~
\mbox{\RM{ if $\beta \rightarrow \pm \infty$ }}. \nonumber
\eey
\beq \label{udefine}
H(k,{y_0 \over x_0},\alpha)={ 2^{1 -{2 \over k}} \over K(k)} \left[
{1 \over \sin 2\alpha }
\left({u^{k \over 2}+ u^{-{k \over 2}} \over 2}\right) ^{-{2 / k}} \right],
~
u \equiv {y_0/x_0 \over \tan\alpha},
\eeq
where $f(\lambda)$ is defined in eq. (\ref{lambdan}), and
the dependence on the bar angle and axis ratio has been collected
in $H(k, y_0/x_0, \alpha)$.
For most interesting combinations of $\beta$ and $n$,
$\xi_1 = (1 \pm 0.1) \sqrt{2} $, and can be approximated within 5\% by
\beq
\xi_1(\beta, n)  \approx {4 \over 3} + {1 \over 6n} \left({\beta \over \beta-3}
\right)^{1 \over 2}.
\eeq
$\xi_0$ also varies less than 30\% from the value for a
Gaussian elliposidal bar ($\xi_0=\pi^{-1/2}$) and can be
approximated within 15\% by
\beq \label{xi0ap}
\xi_0(\beta, n) \approx {2 \over 3} (1- {\ln n \over 2 n})
+ {1 \over 12} \left( {\beta \over \beta-3} \right)^{1 \over 2} (1 - 
{5 \ln n  \over n}).
\eeq

The Gaussian ellipsoidal model studied in \S 2.2 is a special
case with $(\beta, n, k)=(\infty, 2, 2)$. For this case,
$\xi_1 \rightarrow \sqrt{2}, \xi_0=\pi^{-1/2}$.
Another special case is $(\beta, n, k)=(0, \infty, \infty)$,
which describes a rectangular homogeneous bar. The result
for this model is given in Appendix B.

\section{Result For a Homogeneous Rectangular Bar}

For a homogeneous rectangular bar, $(\beta,n,k)=(0,\infty,\infty)$,
\beq
\xi_1(\beta,n) = {4 \over 3},~~\xi_0(\beta,n) = {2 \over 3},~~ H(k,{y_0 \over x_0},\alpha)= {{\mathrm{min}}(u, u^{-1}) \over 4 \sin 2\alpha},
\eeq
where $u$ is defined in eq. (\ref{udefine}). Towards the centre,
\bey
\left< \tau(0,0) \right> &=& {4 \pi G \over c^2}  \left({M \over z_0}\right) 
{1\over 6 \sin 2\alpha} 
{\mathrm{min}}(u, u^{-1}) \\
 & = & 4 \times 10^{-6}~ \left( {M \over 2 \times 10^{10} M_\odot}
{0.5 \mathrm{kpc} \over z_0} \right)
{\mathrm{min}}(u, u^{-1}) {1 \over \sin 2\alpha}
\eey
The optical depth at any field in the $y-z$ plane
\bey 
\left< \tau(y,z) \right> &= & \left< \tau(0,0) \right>,
~~~~~~~~~~\mbox{\RM{if $|y| \le y_{ic}$ and $|z| \le z_0$}} \\\nonumber
			 &=&  \left< \tau(0,0) \right> 
\left( 1-|y|/y_{oc} \over 1 - y_{ic}/y_{oc} \right)^2,
~~\mbox{\RM{if $y_{ic} < |y| \le y_{oc}$ and $|z| \le z_0$}} \\\nonumber
                         &=& 0,
~~~~~~~~~~\mbox{\RM{otherwise,}}
\eey
where
\beq
y_{ic}=|x_0 \sin\alpha-y_0\cos\alpha|,~~y_{oc}=x_0 \sin\alpha+y_0\cos\alpha,
\eeq
are the $y$ coordinates of two of the four corners of the rectangular bar 
in the positive $y$ side.
Note $\left< \tau(y,z) \right>$ is independent of $y$ and $z$ in a region
near the centre.

\section{First Order Correction to the Analytical Results}

Here we derive a first order correction of our analytical formula for
the optical depth, taking into account of the finite distance to the
bulge, and the source luminosity function.  We show that the
first-order correction is zero for the clump giants.

The rigorous expression for the optical depth, as given by eqs. (1) and
(2), can be rewritten as:
\beq \label{taurig}
\left<\tau\right>= {4\pi G \over c^2} {\int_0^{s_{max}} n(D_s) dD_s \int_0^{D_s} dD_d~ \rho(D_d) D 
\over \int_0^{s_{max}}  n(D_s) dD_s },
~~~n(D_s)~ dD_s \propto \rho(D_s) \left( {D_s \over R_0} \right)^{2-2 \beta_s}~ dD_s,
\eeq
where the symbols are the same as in eqs. (1) and (2), and $R_0$ 
is the distance to the Galactic centre.

To the first order of $\left( {a \over R_0} \right)$, where $a$ is
a characteristic length scale of the bar,
\beq
D \approx (x_s-x_l) { R_0+x_l \over R_0 +x_s} \approx (x_s-x_l) - {(x_s-x_l)^2 \over R_0},
\eeq
and
\beq
n(D_s)~ dD_s \propto \rho(D_s)~ dD_s  \left(1 + {x_s \over R_0} \right)^{2-2\beta_s}
\propto \rho(D_s)~ dD_s  \left[1 + (2-2\beta_s) {x_s \over R_0} \right].
\eeq
Changing the lens and source coordinates to the $(x,y,z)$ frame, and
assuming the density of source or lens is zero at the large radius,  we have
\beq\label{tauexpand}
\left< \tau \right> \approx \left< \tau_{1} (y,z) \right> = {4 \pi G \over c^2} 
{I_2 + I_3  \over I_0 + I_1 },
\eeq
where 
\begin{eqnarray}
I_0 &= &  \int_{-\infty}^{+\infty}dx_s ~ \rho(x_s, y, z), \\ 
I_2 &= & \int_{-\infty}^{+\infty}dx_s ~ \rho(x_s, y, z) ~\int_{-\infty}^{x_s} \rho(x_l, y,z) ~ (x_s-x_l)~ dx_l,
\end{eqnarray}
and
\begin{eqnarray}\label{I1first}
I_1 &= & 2(1-\beta_s)\int_{-\infty}^{+\infty}dx_s ~ {x_s \over R_0} \rho(x_s, y, z), \\ \label{I1second}
    &= & -2(1-\beta_s)\int_{-\infty}^{+\infty}dx_s ~ {x_s \over R_0} \rho(x_s, -y, z), \\ \label{I1}
    &= &(1-\beta_s) \int_{-\infty}^{+\infty}dx_s ~ {x_s \over R_0} \left[\rho(x_s, y, z)-\rho(x_s, -y, z)\right], \\ 
    &= & 0,~~ \mbox{\RM{if $y=0$ or $\beta_s=1$,}} 
\end{eqnarray}
\begin{eqnarray} \label{I3first}
I_3 &= & \int_{-\infty}^\infty dx_s \int_{-\infty}^{x_s} dx_l ~\rho_{+y,+y}
~ \left[ -{(x_s-x_l)^2 \over R_0} +2(1-\beta_s) {x_s(x_s-x_l) \over R_0} \right] \\ \label{I3second}
    &= & \int_{-\infty}^\infty dx_s \int_{-\infty}^{x_s} dx_l ~\rho_{-y,-y}
~ \left[ -{(x_s-x_l)^2 \over R_0} +2(1-\beta_s) {x_l(x_l-x_s) \over R_0} \right] \\ \label{I3}
    &=& (1-\beta_s) 
\int_{-\infty}^\infty dx_s \int_{-\infty}^{x_s} dx_l
~\rho_{\mathrm{odd}} \left[ {x_s^2-x_l^2 \over 2R_0} \right]
\\ \nonumber &\mbox{ } &
~~~~~-\beta_s
\int_{-\infty}^\infty dx_s \int_{-\infty}^{x_s} dx_l
~\rho_{\mathrm{even}}
  \left[ {(x_s-x_l)^2 \over 2R_0} \right]  \\ \label{I3minor}
& = & -\beta_s
\int_{-\infty}^\infty dx_s \int_{-\infty}^{x_s} dx_l~
\rho(x_s, 0, z) ~ \rho(x_l, 0,z) ~ \left[ {(x_s-x_l)^2 \over R_0} \right],
~~\mbox{\RM{if $y=0$,}} 
\end{eqnarray}
where
\beq
\rho_{+y,+y}=\rho(x_s, y, z)~\rho(x_l, y,z),~~
\rho_{-y,-y}=\rho(x_s, -y, z)~\rho(x_l, -y,z),
\eeq
\beq
\rho_{\mathrm{even}}=\rho_{+y,+y}+\rho_{-y,-y},~~
\rho_{\mathrm{odd}}=\rho_{+y,+y}-\rho_{-y,-y}.
\eeq

In the above we have used $\left< \tau_{1} (y,z) \right>$ to denote
the first order corrected optical depth.  If we use $\left< \tau_{0}
(y,z) \right>$ to denote the optical depth given by eq.~(\ref{taua}) in
the main text, then $\left< \tau_{1} (y,z) \right>$ reduces to $\left<
\tau_{0} (y,z) \right>$ if the first order correction terms $I_1$
and $I_3$ are set to zero.

Eqs. (\ref{I1second}) and (\ref{I3second}) are derived by
applying a change of the dummy variables
\beq
x_s \rightarrow - x_l, ~~~ x_l \rightarrow - x_s,
\eeq
and the central symmetry of bars
\beq
\rho(-x, y, z)=\rho(x, -y, z),
\eeq
to eqs. (\ref{I1first}) and (\ref{I3first}).
We then obtain 
eqs. (\ref{I1}) and (\ref{I3}) after adding 
eqs. (\ref{I1first}) and (\ref{I1second}), and
eqs. (\ref{I3first}) and (\ref{I3second}),
and then dividing both sides
by two.  Eq. (\ref{I3minor}) is because $\rho_{odd}=0$ on the minor
axis.

A nice result from this is that for red clump stars with
presumably $\beta_s=0$, $I_1=I_3=0$ for any line-of-sight along the
minor axis $y=0$.  Hence $\left< \tau_{1} (0,z) \right>= \left<
\tau_{0} (0,z) \right>$, namely, the optical depth given in the main
text is correct to the first order.  This conclusion is true for any
bar density profile, shape and orientation.

For main sequence sources with $\beta_s>0$, our formula over-estimates
the optical depth and has to be corrected down. For the density profiles
described by eqs. (\ref{rhoxyz}) and (\ref{lambdan}), after some
tedious algebra, the optical depth, corrected to first order, is
given by
\begin{eqnarray}
\left< \tau_{1} (0,z) \right> & = & \left< \tau_{0} (0,z) \right> \left( 1 + {I_3 \over I_2} \right) = \left< \tau_{0} (0,z) \right>
\left[1 - \xi_4 \beta_s {a \over R_0} \left({\rho(0,0,z) \over \rho(0,0,0)}\right)^{-{1 \over \beta}}\right] \\ 
& = & \left< \tau_0(0, z)\right> \left(1-\xi_4
\beta_s {a \over R_0}
\left[1+ {1 \over \beta-3}\left({z
\over z_0}\right)^n \right]^{1 \over n} \right),\label{correct}
\end{eqnarray}
where
\begin{equation} \label{xi4}
{1 \over a^k} ={\cos^{k}\alpha \over x_0^k}+{\sin^{k}\alpha \over y_0^k},
~~\xi_4(\beta,n) \approx 3^{1 \over n}.
\end{equation}
The expression for $\xi_4(\beta,n) \approx 3^{1 \over n}$ is accurate
within 15\% for most bar models, and is exact for homogeneous
($\beta=0$) bar. For Gaussian ellipsoidal bars, $\xi_4=\sqrt{\pi}$.
Other exact values for $\xi_4$ for a few models are given in Table 1.
 From eq. (\ref{correct}), the first-order correction is bigger away from
the plane for power-law models ($\beta > 3$), smaller for truncated
bars ($\beta \le 0$) and independent of $z$ for $\beta \rightarrow \infty$
(e.g., the Gaussian and exponential models).
%
%

We have also checked these formulae with fully
numerical calculations. We find that the above first order corrected
results hold within 5\%-15\% for the full range of models discussed
in the main text, which have $a / R_0 \le {1 \over 4}$.

For readers' convenience, we summarize in the following how to
use our analytical formulae:
\begin{enumerate}
\item{
Find the value of $\xi_0(\beta, n)$. $\xi_0$ for six comibinations
of $(\beta,n)$ are tabulated in Table 1. For combinations not found
in the table, the approximation given by eq. (\ref{xi0ap}) should be
sufficient for most cases. When more accuracy is desired, 
use eqs. (\ref{xi1f}) and (\ref{xi1ratio}).
}
\item{
Use eq. (\ref{udefine}) to calculate the function $H(k, y_0/x_0, \alpha)$.
}
\item{
Now use eq. (\ref{tau0}) to calculate the zeroth-order optical depth.
}
\item{
The first order corrected optical depth can be obtained using
eqs. (\ref{xi4}) and (\ref{correct}).
}
\end{enumerate}

{}


\bsp
\label{lastpage}
\end{document}